\documentclass[natbib]{svjour3}                     
\smartqed  
\usepackage{graphicx}
 \usepackage{aps-bibstyle}  
\newcommand{\aap}{{Astron. Astrophys.}}
\newcommand{\apj}{{Astrophys. J.}}

\begin{document}

\title{Modeling of Photoionized Plasmas}

\author{T. R. Kallman}

\institute{T. Kallman \at
     NASA Goddard Space Flight Center, Greenbelt, MD 20771\\
     Tel.: 301-286-3680
     \email{Timothy.R.Kallman@nasa.gov}
}

\date{Received: date / Accepted: date}

\maketitle

\begin{abstract}
In this paper I review the motivation and current status of 
modeling of plasmas exposed to strong radiation fields, as it applies to the study of cosmic 
X-ray sources.  This includes some of the astrophysical 
issues which can be addressed, the ingredients for the models, the 
current computational tools, the limitations imposed by currently 
available atomic data, and the validity of some of the standard 
assumptions.  I will also discuss ideas for the future:  challenges 
associated with future missions, opportunities presented by improved 
computers, and goals for atomic data collection.
\end{abstract}

\section{Introduction}

The general problem of calculating the reprocessing of ionizing continuum radiation 
from a star or compact object into longer wavelength lines and diffuse continuum
has broad importance in astrophysics.  
Photoionization modeling is generally thought to include any situation in which the dominant 
ionization and excitation mechanism is photons from an external source.  In addition, key ingredients of 
such models are applicable more broadly, to calculating the photoelectric opacity 
and the flux transmitted through any astrophysical gas.
In the context of X-ray astronomy, photoionization is important in sources which contain 
a compact object, notably active galactic nuclei (AGN), X-ray binaries, and cataclysmic variables.

Many of the same principles apply to modeling the properties of diffuse nebulae illuminated 
by UV radiation from stars, i.e. H II regions and planetary nebulae. 
Discussions of model line emission from nebulae date from \cite{Zans27}, but
Seaton and collaborators laid the groundwork for modern numerical modeling by calculating  many atomic quantities of 
importance to nebular modeling  \citep{Seat58, Seat59,  Burg60}.   
The subject has been reviewed extensively by \cite{Oste06}, and others (see references in \cite{Kall01b})
In this volume, many of the relevant ideas have been discussed by \cite{Fost11}.
This paper will not repeat this material, but rather will 
describe the specific developments modeling of photoionized plasmas which are 
relevant to X-ray astronomy.

\section{What is Photoionization}

For the purposes of subsequent discussion, typical  photoionization models calculate 
the ionization, excitation, and heating of cosmic gas by an external source of photons.
The gas is generally assumed to be in a time-steady balance between ionization and recombination, 
and between heating and cooling.  If the gas is optically thin, then a useful scaling 
parameter is the ionization parameter, defined in terms of the ratio of the incident 
ionizing flux to the gas density or pressure.  We adopt the definition $\xi=4 \pi F/n$
where $F$ is the incident energy flux integrated between 1 and 1000 Ry, and $n$ is the gas number 
density \cite{Tart69}; various other definitions are also in use.  
When the gas is isobaric, the appropriate ionization parameter 
is proportional to the ratio ionizing flux/gas pressure.  We adopt the combination $\xi/T$
or $\Xi=\xi/(kTc)$ \cite{Krol81} for this purpose.  This much of photoionization 
modeling is common to today's models for X-ray sources and to more traditional models used for 
nebulae.

Prior to the launch of $Chandra$ and $XMM-Newton$, it was recongnized that objects such 
as AGN and X-ray binaries would have spectral features with diagnostic application when
observed in the X-ray band.  These included $K\alpha$ fluorescence lines from iron 
seen from many objects \citep{Gott95, Asai00}, 
and the warm absobers in the soft X-ray band from AGN \citep{Reyn97}.  
The quantum leap in sensitivity and spectral resolution represented by 
$Chandra$ and $XMM-Newton$ revealed that other new ingredients are needed 
in order to make the models useful for quantitative study of X-ray sources.
This motivated a great deal of work  in the treatment of physical processes 
previously neglected, and in the accuracy and comprehensiveness of atomic data.  
A summary of some of these areas will occupy the remainder of this paper.

\section{Radiation Transport}

Although numerical methods for radiation transfer 
are well establised \citep{Hube01}, their implementation has the potential 
to be very computationally expensive.  Also, radiation transfer depends sensitively 
on the geometrical arrangement of the gas and the sources of illuminating radiation, so 
it is difficult to make a calculation which has wide applicability.  
A calculation must treat the transfer of ionizing continuum into the photoionized gas and
transfer of cooling or reprocessed radiation out of the gas.
For the purposes of X-ray astronomy, it is also 
useful to calculate the entire synthetic spectrum produced by 
the photoionized plasma.  This can then be used for direct fitting to data
using tools such as {\sc xspec} \footnote[1]{http://heasarc.gsfc.nasa.gov/docs/xanadu/xspec}, 
{\sc isis} \footnote[2]{http://space.mit.edu/CXC/isis} or {\sc sherpa} \footnote[3]{http://cxc.harvard.edu/sherpa/}.  

Traditional treatments of radiation transfer make several simplifying assumptions which 
allow for efficient calculation and wider applicability of model results.  These include:
(i) Simplified geometry, such as a plane parallel slab or a spherical shell; (ii) Use 
of single-stream transport of the illuminating radiation; and (iii) Use of escape probability
transport of resonance lines.  We note that X-ray resonance line optical depths are 
typically much smaller than optical or UV line depths, so the effects of line transfer 
approximations are reduced compared with traditional nebular photoionization calculations.

Recent progress in radiation transfer for X-ray photoionization has centered around 
the application of sophisticated accurate transfer treatments for certain limited specialized 
problems.  These include the adoption of the accelerated lambda iteration method in the 
Titan code \citep{Chev06, Roza06, Gonc07}.  Recently, the Monte Carlo method has been used 
by \cite{Sim08, Sim10} to treat the transfer in AGN broad absorption line (BAL) and warm absorber flows.  This latter work
combines radiation transfer with the detailed geometry derived from a numerical hydrodynamic model. 
It has led to important insights into the origin of spectral features such as high velocity lines, 
and X-ray emission lines.  This work illustrates the importance of consideration 
of the detailed geometrical distribution of the gas, and the inherent limitations of 
traditional simplified geometrical assumptions.  

\section{Atomic Data and Comprehensiveness}

The other half of the challenge pf current photoionization modeling is almost 
entirely associated 
with atomic data.  It has long been realized that comprehensive atomic data 
is crucial to accurately calculating the thermal properties of gases hotter than 
$\sim 10^{4}$K \cite{Kwan81}, and that this outweighs considerations such as the 
treatment of radiation transfer for traditional nebular problems.  
This implies consistent treatment of various ionization/excitation processes,
both radiative and collisional, 
and their inverses, including inner shell processes, for all of the ions 
of the $\geq$ 10 most abundant elements.
Similar arguments apply to the calculation of the X-ray opacity of partially 
ionized gases.   Furthermore, the need to calculate synthetic spectra 
introduces a further requirement of accuracy on atomic data:  observed X-ray spectra
can have spectral resolution of $\varepsilon/\Delta\varepsilon\simeq 1000$, and
in order to fit to such data, synthetic spectra must employ wavelengths and 
ionization potentials which are accurate at this level.   Such precision cannot 
be achieved by current atomic physics calculations, and requires dedicated experiments.

A great deal of work has been done to calculate and measure atomic energy levels, cross sections
and transition probabilities for the purposes of astrophysical X-ray photoionization modeling.
A detailed summary would require a dedicated paper; many areas of progress have 
been described by \cite{Beie03, Kall07, Fost11}.  Notable are the measurements 
carried out by the electron beam ion trap (EBIT) and storage rings in Germany and 
Sweden, calculations using the FAC, HULLAC, R-Matrix and Autostructure codes.  
Some of this work will be highlighted in the following sections.

\section{Recent Topics}

\subsection{Iron M-shell Unresolved Transition Array (UTA)}

\begin{figure}[!htb]
\includegraphics*[angle=0, scale=0.4]{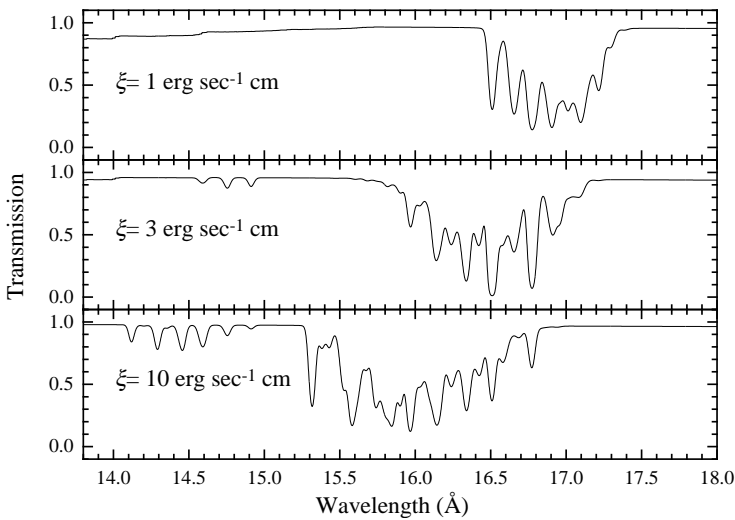}
\caption{\label{beharfig} Iron M shell UTA absorption for 3 different ionization parameters.  From \cite{Beha01}.}
\end{figure}

The $Chandra$ spectrum of the Seyfert 1 galaxy NGC 3783 \citep{Kasp02}, and spectra of 
similar objects, has led to significant revisions of our understanding of these sources.
It has also pointed out several physical effects which were not included in previous photoionization 
models.  One example is the importance of the many line features produced by transitions in iron between the 
levels with principle quantum numbers 2 and 3 (the L and M shells), notably from ions with 10 or more 
electrons, where the  L shell is full.  Previously these lines had only been considered in ions with 
partially filled L shells.  Typical ions have many such lines ($\sim$ 10 -- 100), grouped together 
in wavelength, and the ensemble is called the iron M shell unresolved transition array (UTA). The importance
of these lines was first pointed out by \cite{Beha01}, who also showed that they provide an 
ionization diagnostic for relatively low ionization material.  That is, the centroid shifts from 16 -- 17$\AA$
to  15 -- 16$\AA$ as ionization increases owing to the change from 2p-3d transitions to 2s-3p transitions 
as the 3p shell opens.  This is illustrated in figure \ref{beharfig}, taken 
from \cite{Beha01}.

These features are detected in most Seyfert galaxy warm absorbers, and are important diagnostics of ionization.
In particular, in the case of NGC 3783, they clearly indicate the presence of low ionization iron Fe$^{2+}$ -- Fe$^{8+}$, 
and the relative absence of iron  Fe$^{9+}$ -- Fe$^{16+}$.

\subsection{Fluorescence Lines}

\begin{figure}[!htb]
\includegraphics*[angle=0, scale=0.7]{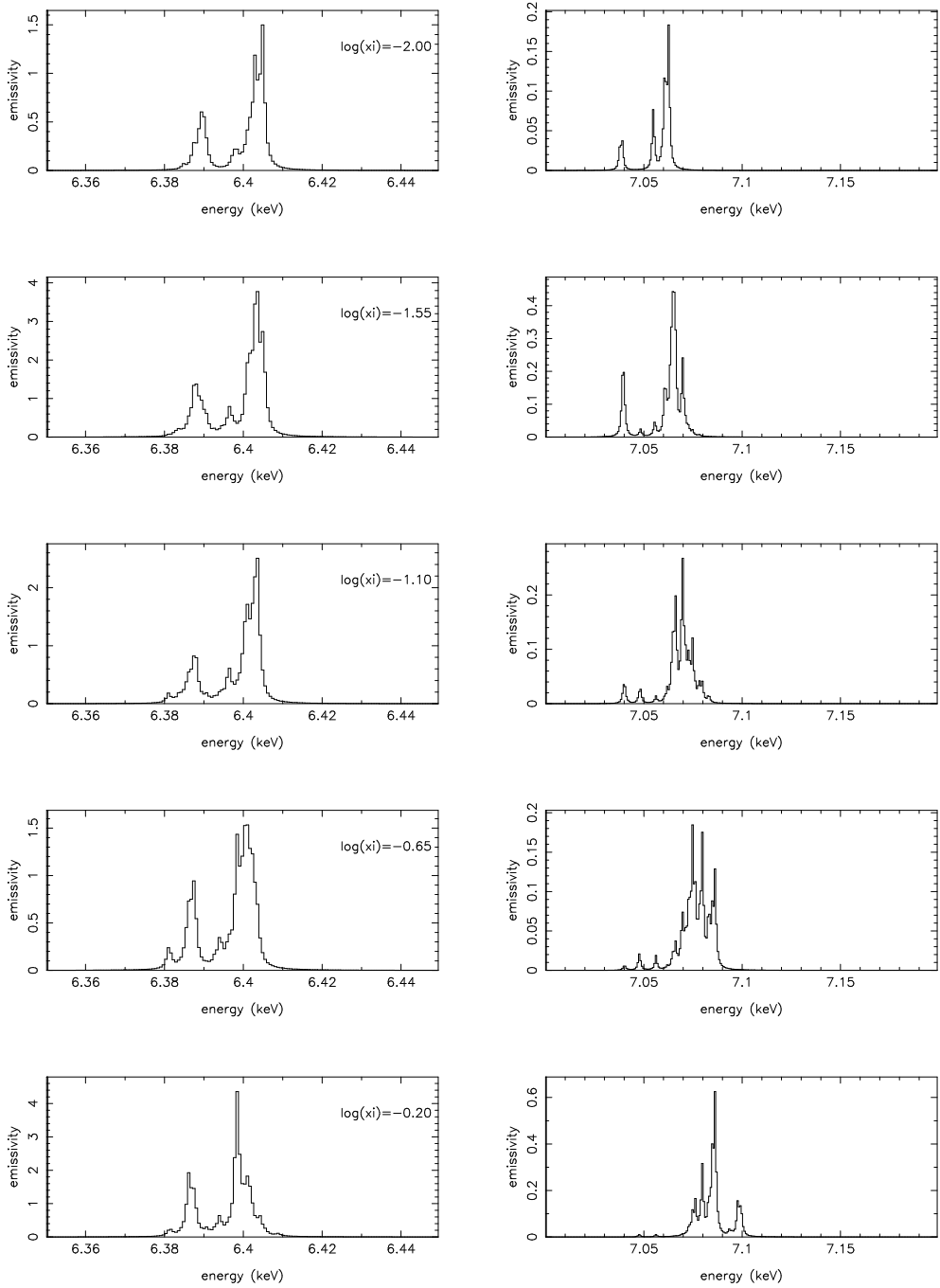}
\caption{\label{kshellspect} Model K shell line spectra of iron for various ionization parameters
(see below for definition).  Emissivity units are arbitrary, but are the same for all panels. \cite{Kall04}.}
\end{figure}

Fluorescence lines can be emitted by gas over a wide range 
of ionization states.  Very useful tables of the properties of these lines 
were provided by \cite{Kaas93}.  
Also, it was pointed out by \cite{Palm03} that the $K\beta/K\alpha$ ratio is a sensitive
diagnostic of ionization, and this has been applied to observed spectra 
by \cite{Yaqo07} and others. Figure \ref{kshellspect} illustrates the variety of iron K emission feature
profiles and strengths that can arise for various ionization states \cite{Kall04}.

\newpage

\subsection{Resonance Scattering}

In traditional nebular modeling it was typically assumed that the gas was spherically 
symmetric around the continuum source and stationary.
If so, resonance scattering of the incident continuum is expected to have negligible effect 
on the spectrum seen by a distant observer.  There is a cancellation between continuum 
photons removed from  and those added to the beam of continuum we observed from the central source. 
In this case, the region of the spectrum emergent from the cloud which includes a resonance line would show neither
a deficit nor an excess at the line energy.  Of course, photoionization followed by recombination
leads to a redistribution of the energy of the photons absorbed by ionization, and so would produce
an excess at the line energy. 

\begin{figure}[!htb]
\includegraphics*[angle=90, scale=0.5]{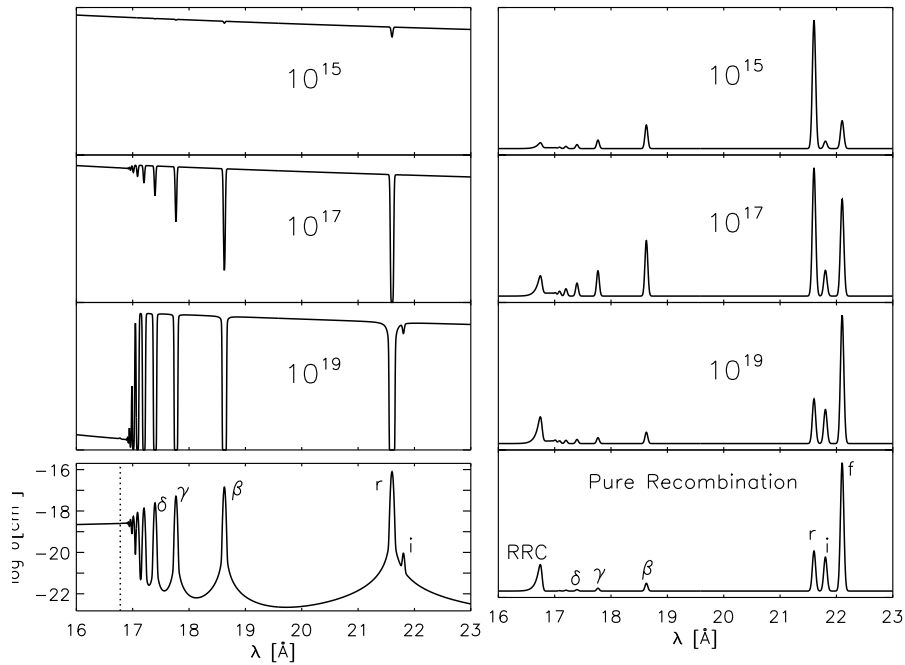}
\caption{\label{scatfig} Scattering-dominated spectrum from \cite{Kink02}.
Left panel shows spectrum seen in transmission, right panel shows
spectrum seen in reflection.}
\end{figure}

It is now recognized that many photoionization-dominated sources observed in X-rays have reprocessing gas which 
is not spherically distributed around the continuum source. 
In the case of a non-spherical scattering region, the cancellation between photons 
scattered out of and into the observed beam is not exact, and thus scattering can
have an effect on the observed spectrum.   The effect depends, crudely, on the relation between the 
column density between the continuum source and infinity averaged over 4$\pi$ steradians, 
as compared with the column density along the observer's line of sight to the continuum source.
If the observer's line of sight traverses more gas than the spherical-average, then resonance scattering 
will result in a net removal  of photons from the observed radiation field and the spectrum will have 
a deficit at the line energy.  This could be termed the 'net absorption' case.
If the observer's line of sight traverses less gas than the spherical-average, then resonance scattering 
will result in a net addition of photons to the observed radiation field and the spectrum will have 
an excess at the line energy.  Such apparent 'net emission' can be confused with emission due to recombination or 
electron impact excitation, though these can be distinguished by differing dependence on atomic 
quantities such as oscillator strengths or collision cross sections.
Furthermore, the resonance scattering cross section is much greater than the background cross 
sections for photoionization, so line are expected to saturate at column densities which 
are small compared to the column densities where the continuum can penetrate.  Recombination emission 
will dominate at higher column densities and scattered emission will dominate at lower column 
densities.   \cite{Kink02}  
have shown that that resonance scattering has a distinct signature in the line ratios and line/continuum ratio
in emitting plasmas.  This is illustrated in figure \ref{scatfig}, which shows the transition from 
a scattering-dominated to a recombination-dominated spectrum as a function of the column density 
of the plasma.  The left column corresponds to what would be observed in the 'net absorption' case, 
and the right column corresponds to 'net emission'.  
At low column densities radiative pumping of resonance lines dominates because 
the cross sections are larger than for photoionization, and the apparent line emission in the 'net emission' case
is due to resonance scattering.  At high column densities the emission is dominated by 
recombination.   

\newpage

\subsection{Dielectronic Recombination}

It was  pointed out by \cite{Netz04} and \cite{Krae04} that the ionization balance needed to fit the K lines of Si in the 
spectrum of NGC3783 was discrepant from that needed to fit the iron UTA lines.  It was suggested 
that this was due to the use of inaccurate dielectronic recombination rates for iron, and it was 
postulated that the rates where were larger by $\sim$10.  This suggestion was confirmed, qualitatively, 
by \cite{Badn06}, who performed distorted wave calculations of the rates and obtained rate coefficients 
which were even greater than those suggested by  \cite{Netz04}.  Experimental 
confirmation was demonstrated for Fe$^{13+}$ \cite{Schm04}, and 
laboratory measurements have been made for ions down to Fe$^{7+}$ \cite{Schi10}.

\begin{figure}[!htb]
\includegraphics*[angle=0, scale=0.4]{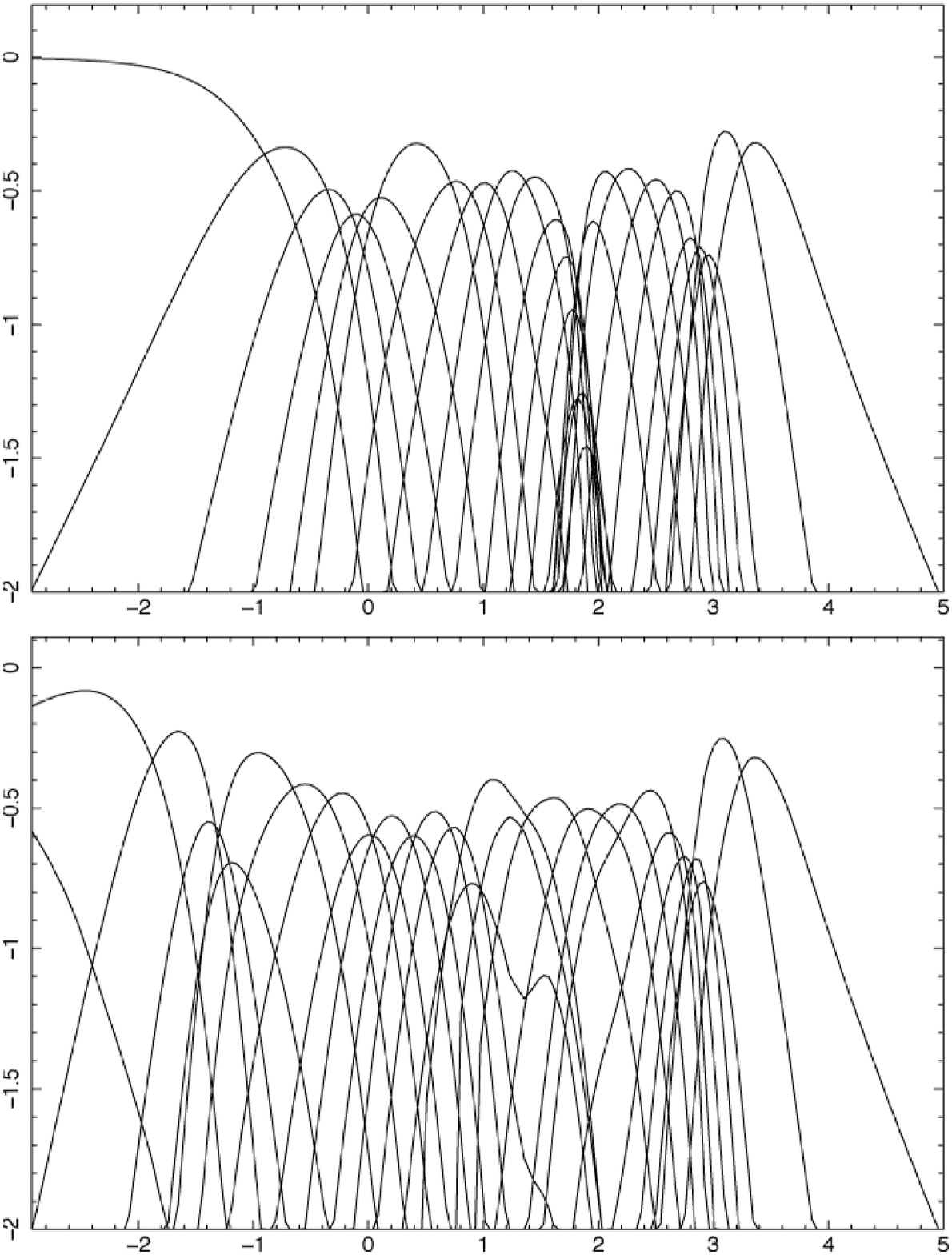}
\caption{\label{ionbalfe}Ionization balance for Fe, ion fractions on the vertical axis
vs. ionization parameter .  New \cite{Badn06} rates for dielectronic 
recombination were used in the calculations shown 
in upper panel, older rates were used in the calculations shown in the lower panel.
Verical axis is log(ion fraction) relative to the total for iron.
Horizontal axis is log$\xi$.  Highest ion fraction is H-like (Fe${25+}$), 
corresponding to right-most curve, and lowest 
is neutral, corresponding to left-most curve, in both panels.}
\end{figure}

\begin{figure}[!htb]
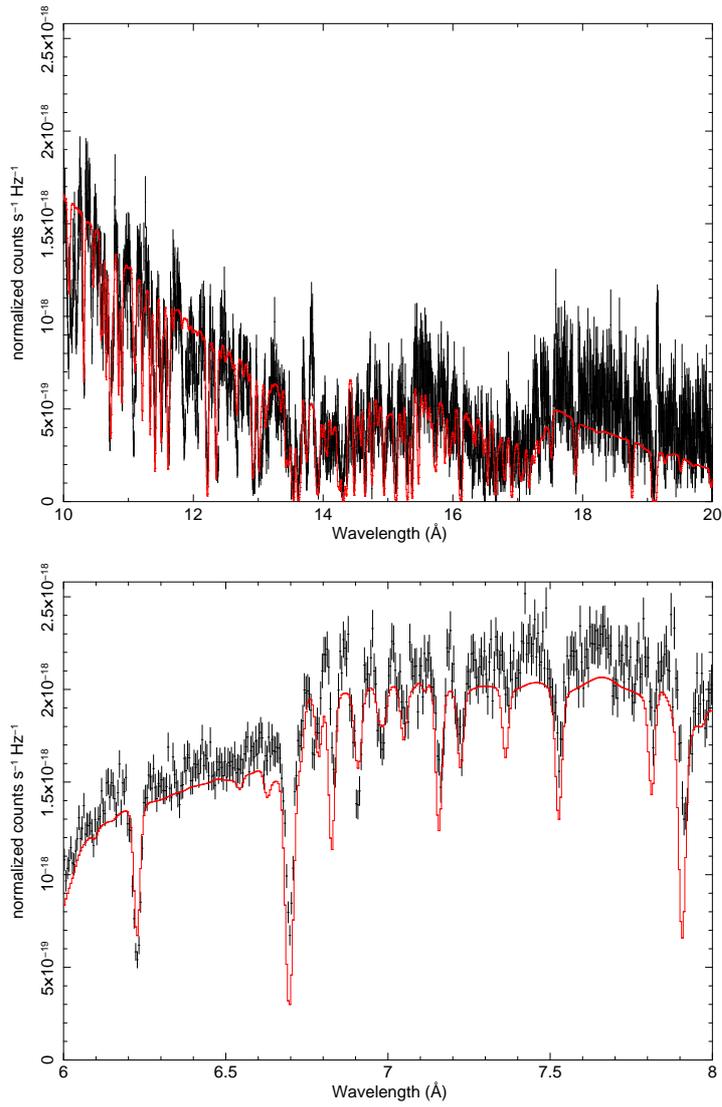

\includegraphics*[angle=270, scale=0.4]{f5.ps}
\includegraphics*[angle=270, scale=0.4]{f6.ps}
\caption{\label{ngc3783fits}Fits to the NGC3783 spectrum using  \cite{Badn06} rates 
for iron dielectronic recombination. Left panel: region including the 
iron UTA near 17 $\AA$.  Right panel: region showing the Si K lines near 6.8-7.2 $\AA$.}
\end{figure}

The implications of these new rates are 
illustrated in figures \ref{ionbalfe} and \ref{ngc3783fits}.  These 
show the iron ionization balance and the fit of the resulting synthetic 
spectrum to the 800 ksec $Chandra$ observation of the Seyfert 1 galaxy NGC 3783
\citep{Kasp02}.
These are shown for two choices of dielectronic recombination rates: the new 
rates, adopting the fits from \cite{Badn06} in the upper panels, 
and the rates which were 
previously used (these are described in \cite{Kall01, Baut01}).  
The fits to the spectrum utilize two components of gas, each with 
a single ionization parameter, and both using the same velocity and line 
broadening.  These are:  v$\simeq$ 800 km/s relative to the galaxy, and 
v$_{turb}\simeq$ 300 km/s.  Figure \ref{ngc3783fits} shows this fit 
using the new rates in the regions containing the iron M shell UTA and the 
Si K lines, showing general agreement.  
The ionization parameters are  log$\xi$=2 and 1
for the new rates.  The ionization balance produced by the older 
rates require log$\xi$=2 and -0.5 
in order to fit the iron UTA.  The Si K lines are missing from the model
since the ionization state of Si is too low to allow K$\alpha$ absorption
at log$\xi$=-0.5.  The fit to the model using the older rates is not shown.
The older rates cannot fit the the iron UTA, 
near 17 $\AA$ and the Si K likes, near 7$\AA$ simultaneously, while the newer ones can.

\newpage

\subsection{Thermal Instability}

\begin{figure}[!htb]
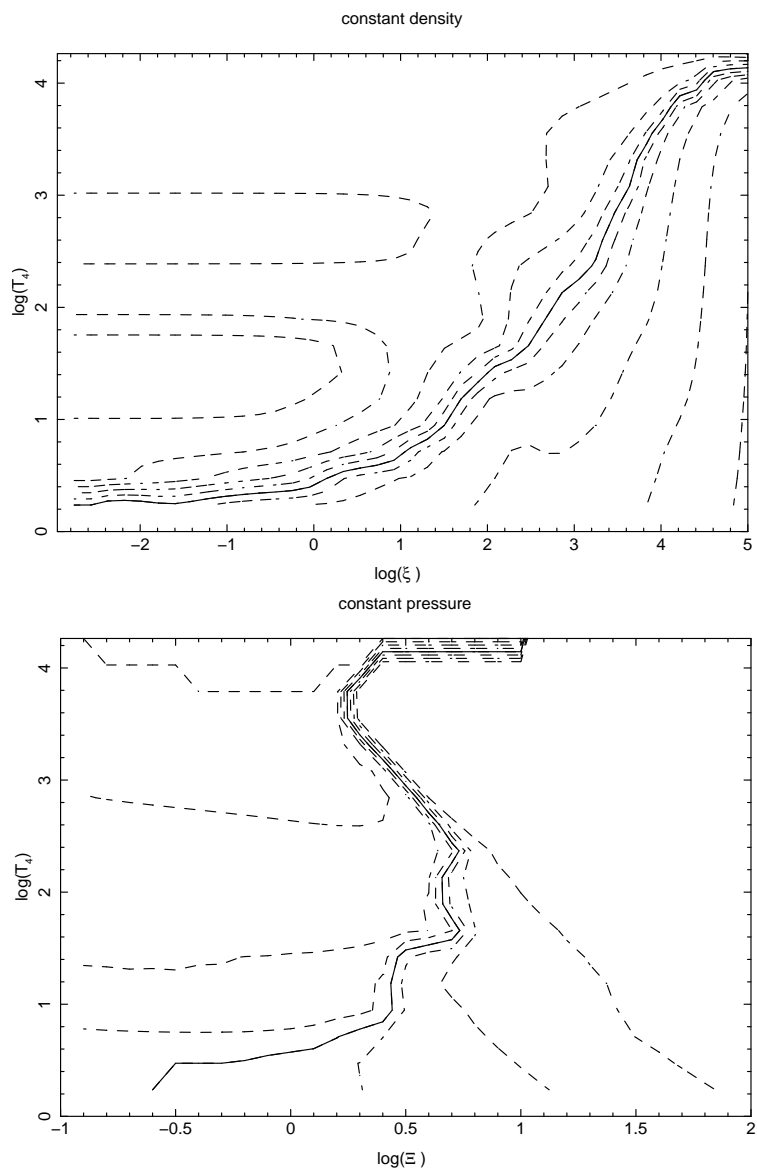

\includegraphics*[angle=270, scale=0.4]{f70a.ps}
\includegraphics*[angle=270, scale=0.4]{f70b.ps}
\caption{\label{txiplot}Contours of constant net cooling-heating in the 
T-$\xi$ plane.  Upper panel:  constant density; lower panel: constant pressure.
Equilibrium is shown as solid curve.}
\end{figure}

A possible explanation for the two phase behavior found in the fits to the 
NGC 3783 spectrum is  thermal instability.  This is  due to 
the properties of the cooling function in the gas which allows for 
two stable and one unstable solutions to the thermal equilibrium equation.
This has been discussed in detail by many authors, notably in the context of 
quasar broad line clouds \cite{Krol81, Math87}.  The physical origin is the 
the fact that the net cooling function has strong temperature dependence in some 
regions of temperature, and weaker dependence in other regions. 
This behavior is illustrated in figure \ref{txiplot}, which shows surfaces 
of constant net cooling (defined below) in the temperature-ionization parameter
plane for a photoionized gas.  Here and in all the examples in this section
we adopt a $\Gamma$=2 power law for the ionizing spectrum illuminating the gas.
That is, the illuminating flux is 
$F_\varepsilon \propto \varepsilon^{1-\Gamma}$ erg cm$^{-2}$ s$^{-1}$ erg$^{-1}$.
This figure shows the equilibrium surface as a solid curve, and illustrates 
the difference between constant density and constant pressure gases.  
In the constant density case there is only a single value of equilibrium 
temperature for a given ionization parameter, while in the constant 
pressure case there is a narrow region of ionization parameter where three
equilibria are possible.  

Thermal instablility is associated with regions where the 
net cooling function $\Lambda(T)$ is a decreasing function of temperature.  
A stable temperature solution is characterized by a cooling function which is an increasing 
function of tempeature;  a positive temperature perturbation about a stable equilibrium 
temperature will lead to increased cooling, which will restore the gas to equilibrium, 
with the corresponding behavior for a negative temperature perturbation.
Perturbations about an unstable temperature will tend
to run away toward higher or lower temperature until a stable region of the cooling curve is reached.
The net cooling function of a photoionized gas is globally increasing over the temperature range from
$\sim$ 1000 K -- 10$^8$K, owing to strong hydrogen cooling at low temperature, 
and strong inverse Compton cooling at high temperature.  
Instability is associated with a local maximum to the function, and so there must always be an odd number 
of thermal equilibrium solutions. 

The instability is more  likely to occur when the gas is isobaric, rather than isochoric.
Isochoric gas heated by a radiation field such as that shown here, i.e. the spectral energy distribution characteristic of AGN,
is predicted to be thermally stable; a radiation field which is flat, or deficient in soft photons, is 
more likely to produce thermal instability.
The presence of the thermal instability  depends on interesting things: the shape of the 
ionizing spectrum (SED) from IR through the X-rays, the atomic rates, elemental abundances,
and (weakly) in the gas density.  This suggests possible diagnostic use.

\begin{figure}[!htb]
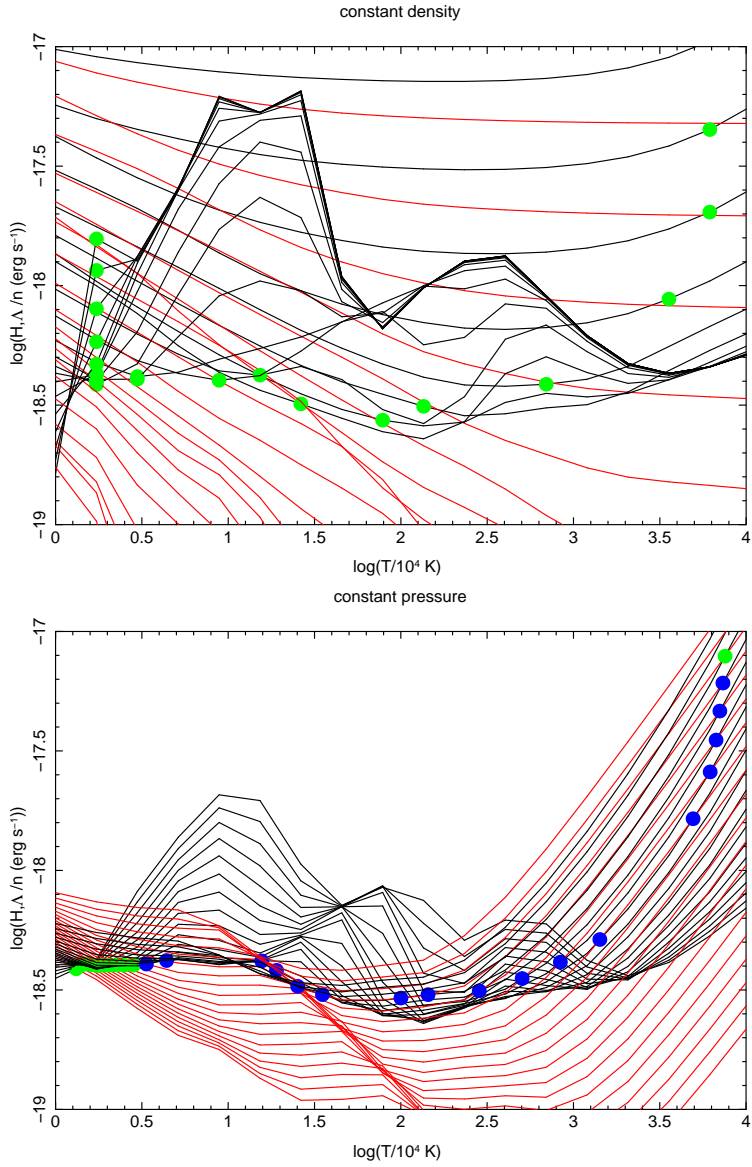

\includegraphics*[angle=270, scale=0.4]{f7a.ps}
\includegraphics*[angle=270, scale=0.4]{f7b.ps}
\caption{\label{hmcrates}Heating (read) and cooling (black) rates vs. temperature.
The upper panel corresponds to constant density gas, and the curves are 
for values of log$\xi$ 0 -- 5. The lower panel corresponds to contant pressure gas 
and the curves are for values of  log($\Xi$) -1 -- 2. 
Vertical axis is log(Rate per hydrogen nuclues) 
in erg  s$^{-1}$.  Equilibrium values are shown as dots, green 
dots for thermally stable solutions.  Blue dots correspond to equilibrium values which 
are either unstable, or which are stable but occur 
for values of $\Xi$ which also have unstable solutions.}
\end{figure}

The same result is displayed in a different way in 
figure \ref{hmcrates}, which plots the heating and cooling functions per hydrogen nucleus vs. temperature for 
various values of the ionization parameter.  The left panel shows this for constant density  gas 
and the right panel is for constant pressure  gas.  In the constant density case the curves 
correspond to various values of the ionization parameter $\xi$, while in the constant pressure case
the curves correspond to values of the ionization parameter $\Xi$.   The figure shows the equilibrium
solutions as colored dots: green dots correspond to  solutions which are unique, i.e. values of 
ionization parameter for which there is only one solution; and blue dots where solutions are not unique.

The existence of 
thermal instability depends on the validity of the assumption of thermal (and ionization) 
equilibrium, and it is important to consider this when using it for quantitative work.
Thermal equilibrium requires that the timescale for heating and cooling, the thermal timescale,
be less than other relevant timescales.  In the case of warm absorbers, which are flowing out
from the AGN center, this includes the gas flow timescale.

The net cooling rate per nucleus in a photoionized gas can be written

\begin{equation}
L_{net}=n\Lambda-H
\label{eq1}
\end{equation}

\noindent where $\Lambda$ is the cooling rate coefficient, and  the heating rate can be written: 

\begin{equation}
H=H_X + H_C
\label{eq2}
\end{equation}

\noindent$H_X$ and  $H_C$ are the photoelectric and Compton heating rates, respectively, and 
can be written:

\begin{equation}
H_C\simeq n \xi \sigma_C \frac{<\varepsilon>-4kT}{m_ec^2}
\label{eq3}
\end{equation}

\noindent where $T$ is the electron kinetic temperature, $<\varepsilon>$ is the mean photon energy, 
$\sigma_C$ is the Compton cross section, $n$ is the gas 
number density.  An approximation to the photoelectric heating rate is \cite{Blon94}:

\begin{equation}
H_X \simeq 1.5 \times 10^{-21} n_{cm3}^2 \xi^{1/4} T_K^{1/2} \left(1-\frac{T}{T_x}\right) {\rm ~erg~s^{-1}}
\label{eq4}
\end{equation}

\noindent where $T_x \simeq 10^6$K is a typical value for power law ionizing spectra and 
$T_K$ is temperature in units of K and $n_{cm3}$ is density in units of cm$^{-3}$ .
An approximation to the cooling rate is \cite{Blon94}:

\begin{equation}
\Lambda \simeq 3.3 \times 10^{-27} T_K^{1/2} + 1.7 \times 10^{-18} \xi^{-1} T_K^{-1/2} e^{-\frac{1.3 \times 10^5K}{T_K}}  {\rm ~erg~cm^3~s^{-1}}
\label{eq5}
\end{equation}

\noindent The first term is due to bremsstrahlung and dominates at temperatures $T_x \leq T \leq T_C$.  
At constant pressure in this range the cooling rate per particle has tempeature dependence 
$n\Lambda \propto T^{-1/2}$, and such gas will be thermally unstable according to the criterion 
of \cite{Fiel65}:  $dL/dT < 0$.

\noindent The cooling time is

\begin{equation}
t_{cool} = \frac{3kT}{L}
\end{equation}

\noindent and using the result from equation \ref{eq5}

\begin{equation}
t_{cool} \simeq 10^{16} n^{-1} T_5^{3/2} \xi_2 \rm{s}
\end{equation}

\noindent where $T_5$ is the temperature in units of $10^5$K and $\xi_2$ is the ionization parameter in units 
of $10^2$ erg cm s$^{-1}$.  The density $n$ is poorly constrained, but likely values are in the range 
$10^4$ -- $10^8$.
 Figure \ref{hmcrates} shows more accurate numerical calculation
of these rates as a function of T and $\xi$.  This shows the difference between constant density and 
constant pressure calculations, and shows the regions of thermal instability.  Constant pressure
is more likely to lead to thermal instability owing to the added inverse dependence of density
on temperature, thereby leading to regions where the net cooling decreases with increasing temperature.

\newpage

\newpage

The timescale for gas to flow outward in the warm absorber is

\begin{equation}
t_{flow} = \frac{R}{v}
\simeq 10^{11} R_{18}v_7^{-1} \rm{s}
\end{equation}

\noindent where R is the typical flow lengthscale, $R_{18}$ is $R/10^{18}$ cm,
and $v$ is the flow velocity.  It is clear from these estimates that the timescales 
 for cooling and flow can be comparable.  Thus, the assumption of thermal equilibrium
must be carefully evaluated before the thermal instability is used as a diagnistic.

\begin{figure}[!htb]
\includegraphics*[angle=0, scale=0.5]{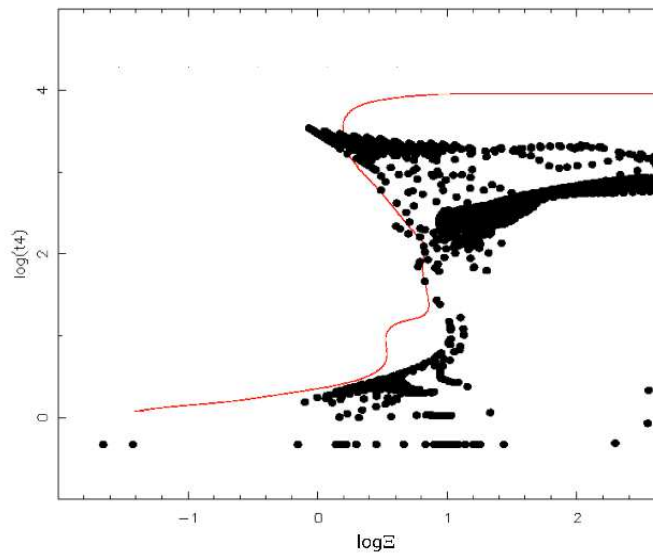}
\caption{\label{xitfig}Locus of points in evaporative torus hydrodynamic model.  This snapshot 
illustrates that many zones are out of thermal equilibrium; time depdendent effects 
are important.  The thermally unstable regions are generally avoided, but not completely.
Equilibrium curve is shown in red.}
\end{figure}

Now we can test this for a more realistic model of the warm absorber.
This is a 2.5 dimensional (3 dimensional axisymmetric) hydrodynamic calculation of the evaporation from the torus responsible 
for the obscuration in Seyfert 2 galaxies.
In this model the torus is heated by a non-thermal $\Gamma$=2 power law from the black hole.
The warm absorber is formed as gas is evaporated and flows out (radiative driving is included)
The thermodynamics of X-ray heating and radiative cooling is included, and the dynamics
are calculated as pure hydrodynamics, no magnetohydrodynamic effects are included.
The synthetic spectrum is also calculated from these simulations.  These have been published 
in \cite{Doro08, Doro09}.  It is interesting to consider what happens in the 
T vs $\xi/T$ plane in such a model.  This is illustrated in figure \ref{xitfig}, 
which shows the loci of $T$ vs $\xi/T$  for such a simulation.  The effect of thermal instability 
is seen for temperatures near $10^{5.5}$K, where there are fewer points.

This  model shows that the thermal properties of warm absorber gas are 
considerably more complicated than most current models.  Departures from 
thermal equilibrium are important.  At the highest temperatures adiabatic cooling is important.
This is a snapshot  at one particular time in the simulation; the loci of points are not static, 
but are constantly moving as material is evaporated from the torus, moves outward and cools.
It illustrates that many zones are out of thermal equilibrium; time dependent effects 
are important.  The thermally unstable regions are generally avoided, but not completely.
Thus, simple stability models provide a very approximate guide for where the gas ends up.  
They may overestime the ionization parameter, since non-equilibrium gas 
may be at a lower ionization parameter than would be inferred from fitting to equilibrium 
models.   We should not be surprised to see gas in unstable regions.
Furthermore, the appearance varies on flow timescale;
 the same model may look different when viewed in many different objects.

\section{The Future}

Fitting models of photoionized plasmas to $Chandra$ and $XMM-Newton$ spectra provides insights to the 
nature of warm absorbers and related structures: their degree of ionization, density, location, composition
and kinematics.  However, the models are likely still incomplete in important ways.  For example,
X-ray grating spectra with good statistics seldom give truly statistically acceptable fits to standard models.
Typically, $\chi^2$ per degree of freedom is $\sim$2 or greater.  Possible reasons include:
missing lines in the atomic database, incorrect treatment of line broadening, incorrect ionization 
balance, overly idealized assumptions (such as ionization equilibrium), inaccurate 
treatments of radiative transfer or geometrical effects.   In addition to thinking about these 
things, modelers need to prepare for the next generation of X-ray instruments, which will likely
have improved sensitivity in the iron K energy band.  This will allow quantitative 
study of lines from trace elements such as  Cr and Mn.
Another frontier is low ionization material;  these instruments may detect 
inner shell fluorescence from many elements with $Z> 10$.   Time dependent effects 
deserve more exploration, as do more user friendly general purpose radiative transfer models.
Photoionization modelers will be helped by laboratory measurements and atomic theory 
if this results in more accurate and comprehensive line wavelengths.
The quest for good fits also depends on accurate instrumental calibration, 
including the response to narrow features, and accurate calibration of continuum for 
accurate subtraction.  Users of photoionization models can provide feedback to modelers and
others in order to detect errors and improve the user interface.


\begin{acknowledgements}
I thank Dr. Ehud Behar and Dr. Daniel Savin for constructive comments, and my collaborators:
M. Bautista, A. Dorodnitsyn, J. Garcia, C. Mendoza, P. Palmeri, P. Quinet, 
and M. Witthoeft.
\end{acknowledgements}


\end{document}